\documentclass[doublecol]{epl2}
\usepackage{amstext,amsmath,amssymb}
\usepackage{makeidx}
\newcommand{\sm}{\scriptscriptstyle}
\newcommand{\avl}{\big\langle}
\newcommand{\avr}{\big\rangle}
\newcommand{\be}{\beta}
\newcommand{\al}{\alpha}

\newcommand{\dm}{\rho}

\newcommand{\p}{\prime}
\newcommand{\drom}{\mathrm{d}}
\newcommand{\iu}{i\,}

\newcommand{\sd}{\mathrm{J}}

\newcommand{\tf}{\Delta}

\newcommand{\w}{\omega}
\newcommand{\Wre}{W^{\al\prime}}

\title{Spin rectification in thermally driven \textit{XXZ} spin chain via the spin-Seebeck effect}

\author{Juzar~Thingna$^{1,2}$\footnote{E-mail address: juzar.thingna@physik.uni-augsburg.de} \and Jian-Sheng~Wang$^1$}
\shortauthor{Thingna~J. and Wang~J.-S.}
\institute{
  \inst{1} Department of Physics and Center for Computational Science and Engineering, National University of Singapore, Singapore 117542, Republic of Singapore\\
  \inst{2} Institut f{\"u}r Physik Universit{\"a}t Augsburg, D-86135 Augsburg, Germany\\
}

\date{5 November 2013}
\abstract{ We study the phenomenon of spin-current rectification in a one-dimensional $XXZ$ spin chain in the presence of a thermal drive. In our model a \emph{pure} spin current is generated by a temperature difference between two harmonic heat baths which in turn creates a spin voltage via the spin-Seebeck effect. Along with a local spin-current operator definition and the nonequilibrium modified Redfield solution we study the spin-rectification ratio as a function of system size and external magnetic field. Intriguing effects are observed at low temperatures such as oscillations with system size and high range of tunability with external magnetic field making magnetic insulators, which are well described by the $XXZ$ model, ideal candidates to build spin devices based on rectification.}
\begin{document}
\maketitle
\section{Introduction}
\label{sec:1}
Manipulation of the spin degree of freedom in order to transport information has led to the mushrooming of the vibrant field of Spintronics \cite{Zutic2004}. Till date most studies are focused on the conventional spin conductors \cite{Appelbaum2007, Tombros2007, Kummeth2008, Stern2008} in which spins are associated with the itinerant charge carriers. The unavoidable motion of these charges leads to a high amount of energy dissipation \cite{Hall2006, Ovichinnikov2008, Trauzettel2008}, which is highly undesirable in order to build low-power devices. The low-energy dissipation along with other advantages, like logic operations \cite{Chen2009} and very large magnetic heat conduction \cite{Sologubenko2000, Hess2001} with mean free paths above 1 $\mu m$ \cite{Hlubek2010}, has generated a surge of theoretical \cite{Zotos2004, Heidrich-Meisner2007} and experimental \cite{Uchida2010, Adachi2010} studies of transport in nonitinerant magnetic systems. 

Some of these nonitinerant magnetic insulator systems like CsCoCl$_{3}$, CsCoBr$_{3}$, SrCuO$_{2}$, Sr$_{2}$CuO$_{3}$, and Cs$_{2}$CoCl$_{4}$ can be modelled by the simple one-dimensional $XXZ$ spin chain model (also known as spin-1/2 anisotropic Heisenberg model) \cite{Yoshizawa1981, Nagler1982, Ami1995, Motoyama1996, Kenzelmann2002, Zaliznyak2004}, whose Hamiltonian is given by
\begin{align}
\label{eq:no2.1}
H_\mathrm{\sm{S}} & = \sum_{i=1}^{N-1}\mathcal{J}\left(\sigma_{i}^{x}\sigma_{i+1}^{x} + \sigma_{i}^{y}\sigma_{i+1}^{y} + \Lambda \sigma_{i}^{z}\sigma_{i+1}^{z}\right)- \sum_{i=1}^{N}h\sigma_{i}^{z},
\end{align}
where $\mathcal{J}$ the exchange coupling between the nearest neighbor spins, $\Lambda$ is the $xz$ anisotropy, $h$ is the external magnetic field along the $z$-direction and $\sigma_{i}^{k}$ ($k = x,y,z$) are the spin-1/2 Pauli matrices of the $i$-th spin. 

The simplified description of the magnetic insulators makes them lucrative candidates for theoretical studies, which has led to a plethora of techniques, like the Mazur inequality \cite{Zotos1997, Prosen2011}, Bethe Ansatz \cite{Zotos1999}, quantum Monte Carlo \cite{Alvarez2002, Louis2003}, Luttinger liquid theory \cite{Sirker2009, Hoogdalem2011}, Hartree-Fock type mean-field approach \cite{Heidrich-Meisner2005}, exact diagonalization techniques \cite{Zotos1996, Fabricius1998, Narozhny1998}, density-matrix renormalization group method \cite{Karrasch2012, Karrasch2013} and the master equation approach \cite{Prosen2009, Zindaric2010, Zindaric2011, Popkov2012}, to probe the transport properties in various parameter regimes. Although these techniques have greatly advanced our understanding of transport coefficients, they have failed to shed enough light on the regime far from linear response. Very little is known in this regime, although new fascinating phenomena \cite{Benenti2009} may exist.

In this letter our goal is to study spin-current rectification in the far from linear response regime for finite-sized one-dimensional $XXZ$ model. In order to minimize dissipation losses, we generate pure spin currents \cite{Brataas2002, Nakata2011} via the spin-Seebeck effect \cite{Bauer2012, Adachi2013} and study the effects of system-size and external magnetic field on the spin-rectification ratio. Our results indicate that magnetic ordering plays a vital role in finite sized systems to describe transport properties at low temperatures. In particular a broken magnetic ordering can cause the spin-rectification ratio to be affected. This low temperature regime also exhibits the feature of tunability of the rectification ratio using an external magnetic field. We hope that this understanding of the spin-rectification ratio in thermally driven $XXZ$ model helps build efficient spin-based devices like transistors and diodes using magnetic insulators.
\section{Spin Current}
\label{sec:2}
The model Hamiltonian for magnetic insulators given by eq.~(\ref{eq:no2.1}), being an integrable quantum model \cite{Takahashi1999}, possess a macroscopic number of nontrivial conservation laws (refer~\cite{Zotos1997} and references therein). One of which is the conservation of total spin along the $z$-direction. The spin conservation law permits us to write a lattice continuity equation and hence define a local spin operator as, 
\begin{align}
\label{eq:no2.2}
\frac{\drom\sigma_{i}^{z}}{\drom t} & = j_{\left(i-\sm{1}\right)\rightarrow i} - j_{i 
\rightarrow \left(i-\sm{1}\right)},\\
\label{eq:no2.3}
j_{n\rightarrow m} & = 
2\mathcal{J}\left(\sigma_{n}^{x}\sigma_{m}^{y}-\sigma_{n}^{y}\sigma_{m}^{x}\right),
\end{align}
where $j_{n \rightarrow m}$ is the $(n,m)$-th element of the local spin operator $j$, 
indicating the flow of spin from site $n$ to site $m$. Throughout this work $\mathcal{J}$ will be used as the scale for all physical quantities and $\hbar$, $k_{\mathrm{\sm{B}}}$ will be set to unity.

Typically, in order to evaluate the spin current one uses the local spin current operator $j$, eq.~(\ref{eq:no2.3}), along with the reduced density matrix $\dm$, obtained by the master equation approach, to obtain the average spin current $ j_{\mathrm{s}} = \avl j \avr = \mathrm{Tr}\left(\dm j\right)$. Till date, the reduced density matrix has always been calculated using the Lindblad formulation \cite{Lindblad1976} where an asymmetry is introduced in the Lindblad operators of the two leads which drives a spin current in the system \cite{Prosen2009, Zindaric2010, Zindaric2011, Popkov2012}. The driving parameter and the Lindblad operators are \emph{phenomenologically} justified as representing a spin-chemical potential and magnetic leads, but the actual microscopic form of the Hamiltonian from which these operators arise is intractable. Temperature in the Lindblad formulation is also undefined and since the Lindblad operators for the equilibrium case give a uniform probability distribution it is generally assumed that it corresponds to a system at infinite temperature.

In order to avoid such phenomenological problems we treat the baths as a set of harmonic oscillators and thus the total Hamiltonian is given by,
\begin{align}
\label{eq:no2.4}
H_{\mathrm{\sm{tot}}} &= H_{\mathrm{\sm{S}}} + H_{\mathrm{\sm{L}}} + H_{\mathrm{\sm{R}}},\\
\label{eq:no2.5}
H_{\al} &= \sum_{n=1}^{\infty} \frac{p_{n,\al}^{2}}{2m_{n,\al}}+\frac{m_{n,\al}\,\w_{n,\al}^{2}}{2}\left(q_{n,\al}-\frac{c_{n,\al}\,S^{\al}}{m_{n,\al}\,\w_{n,\al}^{2}}\right)^{2},
\end{align}
where $\al = \mathrm{L,R}$ and $H_{\mathrm{\al}}$ comprises of the bath and system-bath interaction Hamiltonian. Most common choice of the spin-boson interaction $S^{\al}$ is $\sigma^{z}$ coupled to the collective position operator $\sum_{n}c_{n,\al}q_{n,\al}$ of the bath \cite{Leggett1987}. This type of $\sigma^{z}$ coupling alone cannot induce currents since it allows the bath to only change the energy of the spins. In order to generate spin current the baths should be allowed to make the spins flips, which can be accomplished by the $\sigma^{x}$ or $\sigma^{y}$ coupling. In this work we make one particular choice in the above Zwanzig-Caldeira-Leggett model by choosing $S^{\sm{\mathrm{L}}} = \sigma_{1}^{x}$ and $S^{\sm{\mathrm{R}}} = \sigma_{\sm{N}}^{x}$. A keen reader might notice the presence of a counter-term $\propto (S^{\al})^{2}$ which is sometimes included \cite{Wilhelm2004} or dropped \cite{Segal2005} from such spin-boson interactions. In this work we keep this counter-term since it is merely a matter of convention.

Since the baths have no magnetization the concept of spin-chemical potential is inapplicable and hence such a driving is not possible. Fortunately, the spin-Seebeck effect provides us with an alternate thermal driving field in such nonmagnetized harmonic chain baths. Analogous to the Seebeck effect \cite{Ashcroft1976} studied in electronic transport, the spin-Seebeck effect \cite{Bauer2012, Adachi2013} generates a spin voltage in presence of a temperature difference which in turn generates a spin current in the system. Thus, without resorting to the \emph{ad-hoc} Lindblad formulation we can still calculate the spin currents in the system with thermal driving field. This also justifies our use of the local current operator since in our unique model the alternate definition of spin current, i.e., $\drom\sigma_{\mathrm{\sm{L}}}^{z}/\drom t$ is inapplicable.

Now we focus on the evaluation of the reduced density matrix (RDM) to calculate the spin current. Various types of currents like heat, electronic and also spin at the lowest order of perturbation are second order in the system-bath coupling. In our specific model described above, the local spin-current operator $j$, eq.~(\ref{eq:no2.3}), is independent of system-bath coupling strength and hence in order to accurately capture the nonequilibrium effects it is essential to obtain the RDM correct up to second order in coupling. Since all perturbative master equations are inaccurate at the second order in the steady state \cite{Mori2008, Fleming2011, Thingna2012}, we resort to the novel nonequilibrium modified Redfield solution (NMRS) \cite{Thingna2013} which accurately captures all second order steady state effects. According to the NMRS approach the diagonal $0$-th order RDM in the energy eigen basis of the system Hamiltonian is given by the solution of
\begin{align}
\label{eq:no2.6}
&\sum_{\al ,k}\left(S_{nk}^{\al}S_{kn}^{\al}\Wre_{nk}-\delta_{n,k}\sum_{l}S_{nl}^{\al}S_{lk}^{\al}\Wre_{lk}\right)\dm_{kk}^{(0)} = 0,
\end{align}
where 
\begin{align}
\label{eq:no2.7}
\Wre_{kl} &= \sd^{\al}(\tf_{kl})n^{\al}(\tf_{kl}),
\end{align}
\begin{figure}
\begin{center}
\includegraphics[scale=0.35]{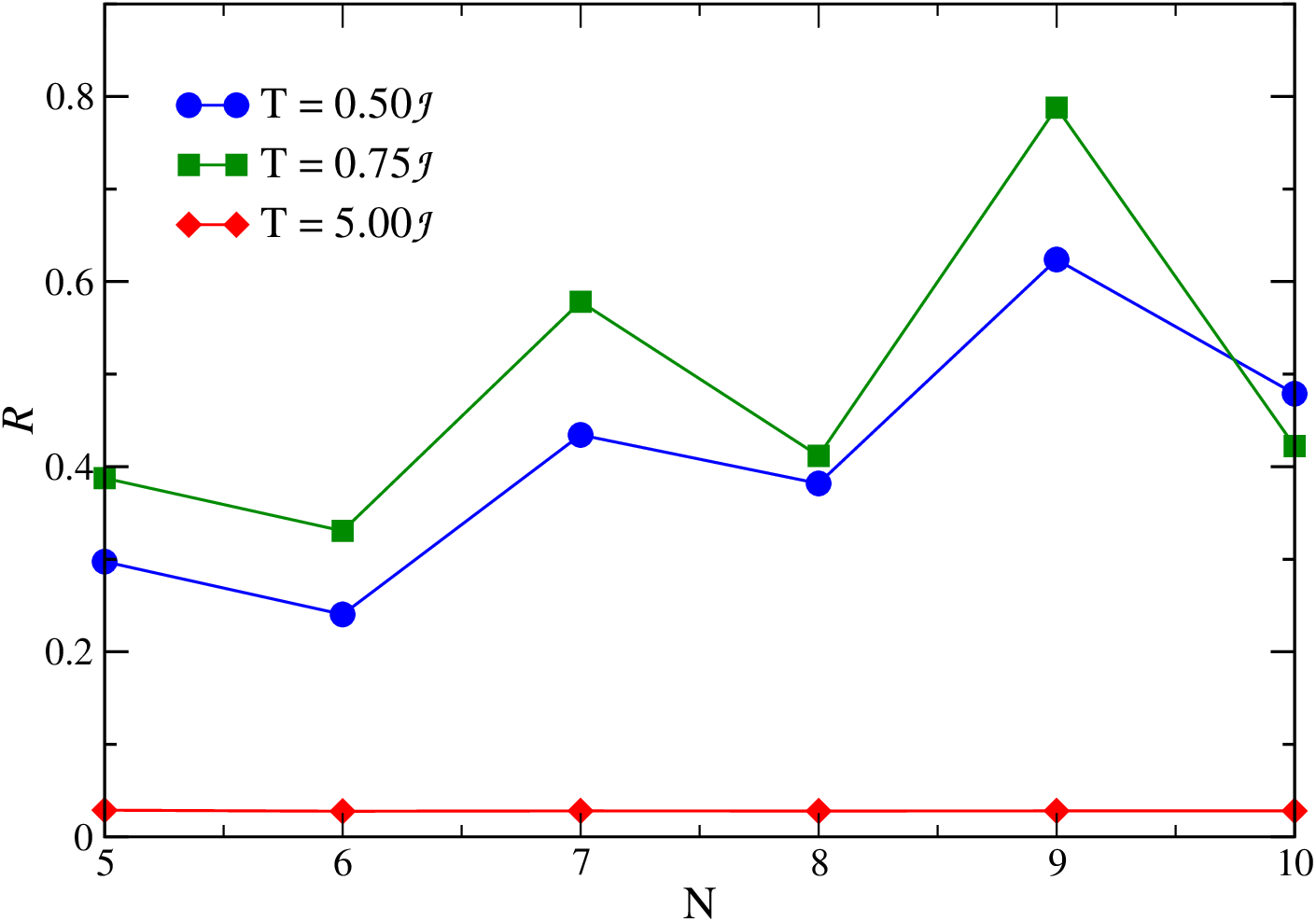}
\end{center}
\caption{(Color Online) Plot of rectification ratio $R$ as a function of system size $N$ for different average temperatures $T = (T_{\mathrm{\sm{L}}}+T_{\mathrm{\sm{R}}})/2$. The parameters used for the simulations are: $\Lambda = 1.5$, $h = 0.5\mathcal{J}$, $\Delta T = T_{\mathrm{\sm{L}}}-T_{\mathrm{\sm{R}}} = 0.5\mathcal{J}$, $\eta^{\mathrm{\sm{L}}} = 0.019\sqrt{\mathcal{J}}$, $\eta^{\mathrm{\sm{R}}} = 0.001\sqrt{\mathcal{J}}$, and $\w_{\mathrm{\sm{D}}} = 10\mathcal{J}$.}
\label{fig:1}
\end{figure}
$\tf_{kl} = E_{k}-E_{l}$ is the difference between two system eigenvalues, $\sd^{\al}(\w) = \pi\sum_{n}\left(c_{n,\al}^{2}/2m_{n,\al}\w_{n,\al}\right)\delta(\w-\w_{n,\al})$ is the spectral density which defines all bath properties and $n^{\al}(\w) = \left[\mathrm{exp}(\be^{\al}\w)-1\right]^{-1}$ is the Bose-Einstein distribution function. In consonance with the NMRS approach the vital $2$-nd order elements of the RDM take the form,
\begin{align}
\label{eq:no2.8}
\dm_{nm}^{(2)} & = \frac{1}{\iu\tf_{nm}} \sum_{\al ,k} S_{nk}^{\al}S_{km}^{\al} \biggl[ \Bigl(W_{nk}^{\al}+W_{mk}^{\al*}\Bigr)\dm_{kk}^{(0)} \biggr. \nonumber\\
& \biggl.-W_{kn}^{\al*}\dm_{nn}^{(0)}-W_{km}^{\al}\dm_{mm}^{(0)}\biggr], ~~~~(n \neq m)\\
\label{eq:no2.9}
\dm_{nn}^{(2)} & = \sum_{\al ,k} S_{nk}^{\al}S_{kn}^{\al} \left[V_{nk}^{\al\p\p}\dm_{kk}^{(0)}-V_{kn}^{\al\p\p}\dm_{nn}^{(0)}+W_{kn}^{\al\p\p}\frac{\partial\dm_{nn}^{(0)}}{\partial E_{n}}\right] \nonumber \\
&-\dm_{nn}^{(0)}\sum_{\al ,k,l}S_{lk}^{\al}S_{kl}^{\al}W_{kl}^{\al\p\p}\frac{\partial\dm_{ll}^{(0)}}{\partial E_{l}},
\end{align}
where 
\begin{align}
\label{eq:no2.10}
W_{kl}^{\al\p\p} &= \mathrm{P}\int_{-\infty}^{+\infty}\frac{\drom\w}{\pi}\frac{\sd^{\al}(\w)n^{\al}(\w)}{\w-\tf_{kl}}+\int_{0}^{\infty}\frac{\drom\w}{\pi}\frac{\sd^{\al}(\w)}{\w},\\
\label{eq:no2.11}
W_{kl}^{\al} &= \Wre_{kl}+\iu W_{kl}^{\al\p\p}\\
\label{eq:no2.12}
V_{kl}^{\al\p\p} &= \frac{\partial W_{kl}^{\al\p\p}}{\partial\tf_{kl}},\\
\label{eq:no2.13}
\frac{\partial\dm_{nn}^{(0)}}{\partial E_{n}} & = \frac{\sum_{\al ,\substack{k\ne n}}S_{nk}^{\al}S_{kn}^{\al}\left(V_{nk}^{\al\p}\dm_{kk}^{(0)}+V_{kn}^{\al\p}\dm_{nn}^{(0)}\right)}{\sum_{\al ,\substack{k\ne n}}S_{nk}^{\al}S_{kn}^{\al}\Wre_{kn}},\\
\label{eq:no2.14}
V_{kl}^{\al\p} &= \frac{\partial W_{kl}^{\al\p}}{\partial\tf_{kl}}.
\end{align}

The $2$-nd order accuracy in the RDM described above is a pivotal feature, since the entire contribution to the spin current arises solely from these $2$-nd order elements. In other words, if one uses the $0$-th order RDM $\dm^{(0)}$ the spin current is \emph{exactly} zero. Besides the accuracy, another salient feature which makes the NMRS favourable to treat spin chains is the improved computational efficiency \cite{Thingna2013}, as compared to other quantum master equations. Thus, the local-current operator given by eq.~(\ref{eq:no2.3}) and the nonequilibrium modified Redfield solution described above give us a computationally easy and accurate prescription to calculate the spin current, arising due to thermal driving, in $XXZ$ spin chain.
\section{Spin Rectification}
\label{sec:3}
\begin{figure}
\begin{center}
\includegraphics[scale=0.3]{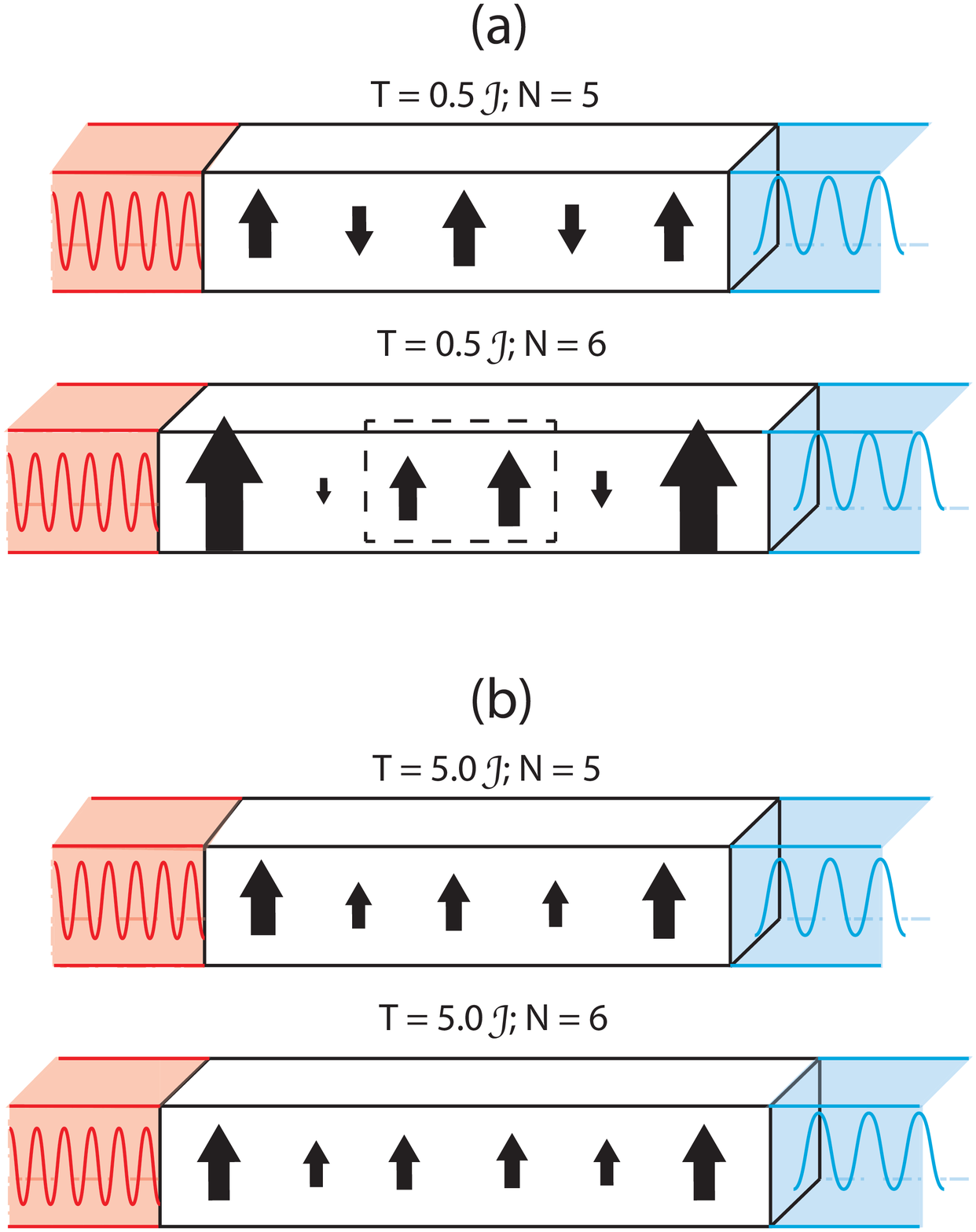}
\end{center}
\caption{(Color Online) Illustration of the $\avl\sigma_{i}^{z}\avr$ at each site $i$ for a 5 site (top) and 6 site (bottom) system. Figures (a) and (b) and for low and high average temperatures $T = (T_{\mathrm{\sm{L}}}+T_{\mathrm{\sm{R}}})/2$ respectively. The length of the arrow is proportional to the value of $\avl\sigma_{i}^{z}\avr$ and the lengths in figs. (a) [bottom] and (b) have been scaled up by a factor of 5. The other parameters used for the simulation are: $\Lambda = 1.5$, $h = 0.5\mathcal{J}$, $\Delta T = T_{\mathrm{\sm{L}}}-T_{\mathrm{\sm{R}}} = 0.5\mathcal{J}$, $\eta^{\mathrm{\sm{L}}} = 0.019\sqrt{\mathcal{J}}$, $\eta^{\mathrm{\sm{R}}} = 0.001\sqrt{\mathcal{J}}$, and $\w_{\mathrm{\sm{D}}} = 10\mathcal{J}$.}
\label{fig:2}
\end{figure}
We now explore spin-current rectification in the far from linear response regime for our one-dimensional $XXZ$ spin chain model. Even though the rectification effect exists in the linear response regime the effect becomes significantly weaker\cite{ThingnaPhD2013}. This makes it difficult to study the variations in rectification as a function of the tunable parameters especially in experimental set-ups, where a high signal-to-noise ratio is required. Hence, in our theoretical calculations we choose the far from linear response regime, which is more accessible to experimentalists, keeping a significant temperature difference between the two leads. Hoogdalem and Loss \cite{Hoogdalem2011} have studied rectification in bulk sized $XXZ$ spin chain, under the assumption of ballistic transport, using a wide variety of perturbative techniques, with different regions of validity. Their work encompasses: Renormalization group, wherein they consider only the low energy excitation (equivalent to low temperature); Luttinger liquid formulation, wherein they consider $\Lambda \ll 1$ and $h \ll \mathcal{J}$, to essentially treat the system as ballistic and spin-wave formulation, wherein $\mathcal{J} < 0$, $\Lambda > 1$, and temperature is low. Despite their extensive efforts their techniques fail to capture the anti-ferromagnetic \cite{Mikeska2004} ($\Lambda > 1$ and $\mathcal{J} > 0$) regime for finite sized spin chains. In this regime, at zero external magnetic field, there has been mounting evidence that the spin transport is mainly diffusive \cite{Prosen2009, Zindaric2010, Zindaric2011,Langer2009, Steinigeweg2009}, whereas for $\Lambda < 1$ the system exhibits a ballistic behavior \cite{Prosen2011, Zindaric2011, Shastry1990}. In this section we will focus on rectification in this unexplored anti-ferromagnetic transport regime with $\mathcal{J} > 0$ and $\Lambda > 1$, where our approach can be easily applied.

We begin by construing rectification as the phenomenon in which if we interchange the 
temperature of the two baths the backward current would have a different value as 
compared to the forward current (obtained when the bath temperatures are not 
interchanged). In order to study this effect we define the forward spin current 
$j_{\mathrm{s}}^{+}$ as the current flowing out of the left bath, which is at a 
temperature $T_{\mathrm{\sm{L}}} > T_{\mathrm{\sm{R}}}$, and the backward current 
$j_{\mathrm{s}}^{-}$ as the one flowing out of the right bath when the temperatures of 
the left and right bath are interchanged, i.e., $T_{\mathrm{\sm{R}}}^{\p} 
(=T_{\mathrm{\sm{L}}}) > T_{\mathrm{\sm{L}}}^{\p} (=T_{\mathrm{\sm{R}}})$. Therefore mathematically we can define the forward and backward currents as,
\begin{align}
\label{eq:no3.0}
j_{\mathrm{s}}^{+} &= \sum_{n,m} j_{n\rightarrow m}\rho_{mn}^{+},\\
j_{\mathrm{s}}^{-} &= \sum_{n,m} j_{n\rightarrow m}\rho_{mn}^{-},
\end{align}
where $\rho_{mn}^{+}$ is calculated for $T_{\mathrm{\sm{L}}} > T_{\mathrm{\sm{R}}}$ and $\rho_{mn}^{-}$ is calculated with the primed temperatures, i.e., $T_{\mathrm{\sm{R}}}^{\p} 
(=T_{\mathrm{\sm{L}}}) > T_{\mathrm{\sm{L}}}^{\p} (=T_{\mathrm{\sm{R}}})$ using eqs.~(\ref{eq:no2.6})--(\ref{eq:no2.14}). Given these definitions we can now quantify the rectification effect \cite{Li2012} using a ratio given by,
\begin{align}
\label{eq:no3.1}
R & = \left|\frac{j_{\mathrm{s}}^{+} - j_{\mathrm{s}}^{-}}{j_{\mathrm{s}}^{+} + j_{\mathrm{s}}^{-}}\right|.
\end{align}
Since we will be focusing on the regime $\Lambda > 1$ there is inherent anharmonicity in the system, which is one of the essential ingredients to have rectification. The other important ingredient is asymmetry, which we induce by coupling the system with heat baths having different physical properties. We choose baths of the ohmic type with a Lorntz-Drude cut-off having the spectral density of the form,
\begin{align}
\label{eq:no3.2}
\sd^{\al}(\w)  &=  \frac{\eta^{\al} \w}{1+(\w/\w_{\mathrm{\sm{D}}})^{2}},
\end{align}
where $\eta^{\al}\propto \sum_{n}c_{n,\al}^{2}$ represents the strength of the system-bath coupling for each bath.

Figure~\ref{fig:1} shows the rectification ratio $R$ as a function of the system size $N$. Clearly the rectification ratio $R$ oscillates with system size, which is highly prominent at low temperatures and becomes negligible as the temperature increases. In order to understand the origin of these oscillations we illustrate in fig.~\ref{fig:2} the $\avl\sigma_{i}^{z}\avr$ at each site for a system comprising of 5 and 6 spins at two different average temperatures $T = (T_{\mathrm{\sm{L}}}+T_{\mathrm{\sm{R}}})/2 = 0.5 \mathcal{J}$ and $ = 5.0\mathcal{J}$. Since the two end spins $i=1,N$ are connected to the baths their $\avl\sigma_{\sm{1/N}}^{z}\avr$ values are greatly influenced by the bath temperature and coupling, which causes the $z$-component of the two end spins to be pinned in the same direction. This pinning allows an anti-ferromagnetic ordering only in the case of odd number of spins as shown in the top illustration of fig.~\ref{fig:2}(a). In case of even number of spins [fig.~\ref{fig:2}(a) bottom] the system cannot attain its lowest energy anti-ferromagnetic state due to the presence of a ferromagnetic bond [enclosed in dashed lines in fig.~\ref{fig:2}(a)]. The spin frustration causes the $\avl\sigma_{i}^{z}\avr$ to be lower at each site which in turn affects the rectification ratio $R$. The oscillatory behaviour is also sensitive to the temperature $T$ and if the temperature becomes very high the system does not have any particular ordering [see fig.~\ref{fig:2}(b)], thus drastically reducing the rectification ratio and suppressing the oscillatory behaviour (depicted by the red solid line with diamonds in fig.~\ref{fig:1}). Thus we deduce that the oscillations in the rectification ratio $R$ as a function of system size crucially depends on the ordering of spins in the finite sized spin chain. 
\begin{figure}
\begin{center}
\includegraphics[scale=0.35]{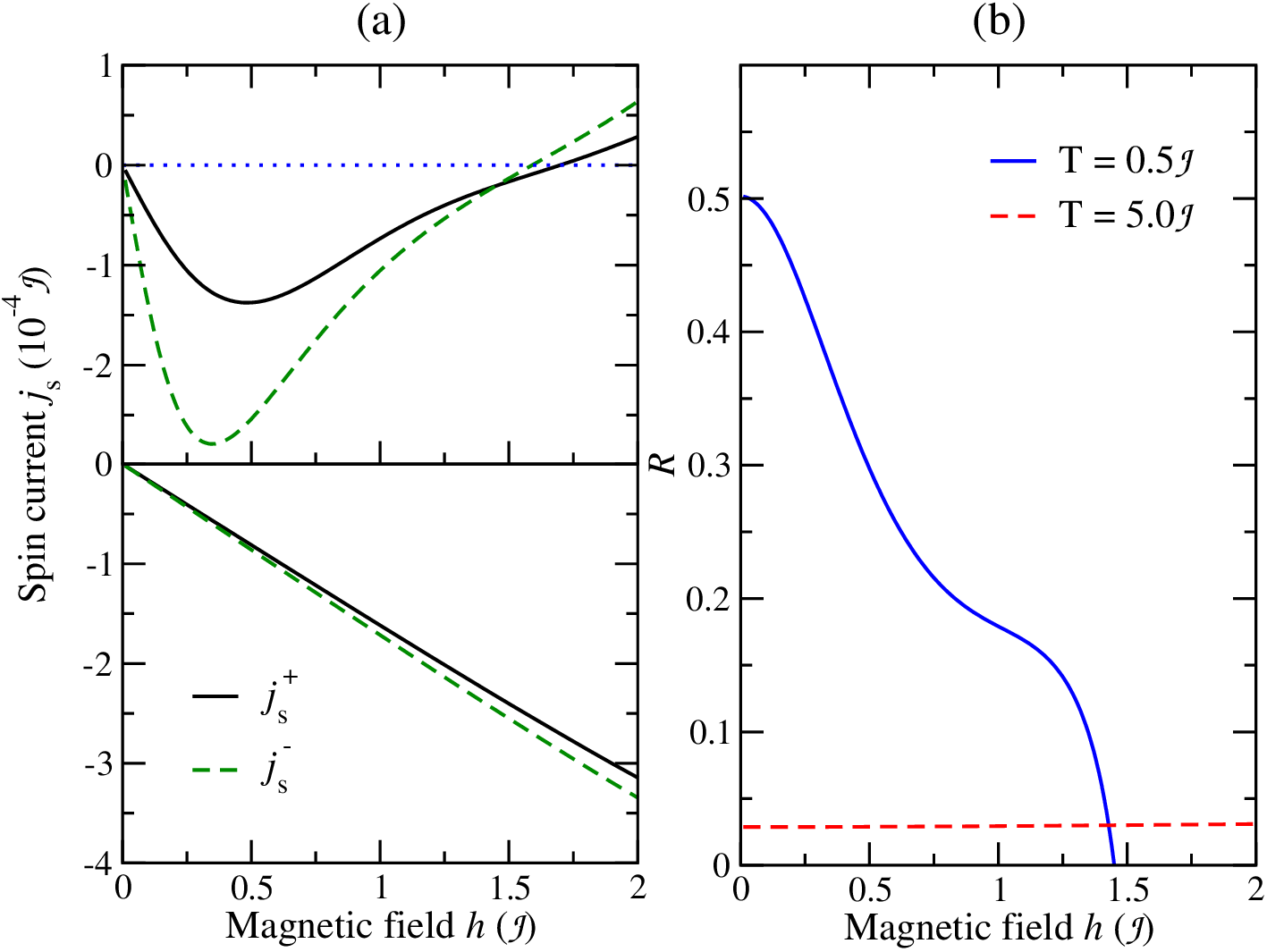}
\end{center}
\caption{(Color Online) The forward $ j_{\mathrm{s}}^{+}$ (black solid) and backward $ j_{\mathrm{s}}^{-}$ (green dashed) currents as a function of the magnetic field $h$ are shown in fig.~(a) for average temperature $T = (T_{\mathrm{\sm{L}}} + T_{\mathrm{\sm{R}}})/2 = 0.5\mathcal{J}$ (top) and $T = 5.0\mathcal{J}$ (bottom). In fig.~(b) we plot the rectification ratio $R$ vs $h$ at average temperature $T = 0.5\mathcal{J}$ (solid blue) and $T = 5.0\mathcal{J}$ (dashed red). The common parameters used are: $N = 5$, $\Lambda = 1.5$, $\Delta T = T_{\mathrm{\sm{L}}}-T_{\mathrm{\sm{R}}}=0.5\mathcal{J}$, $\eta^{\mathrm{\sm{L}}} = 0.019\sqrt{\mathcal{J}}$, $\eta^{\mathrm{\sm{R}}} = 0.001\sqrt{\mathcal{J}}$, and $\w_{\mathrm{\sm{D}}} = 10 \mathcal{J}$.}
\label{fig:3}
\end{figure}

Next in fig.~\ref{fig:3} we show the effects of the external magnetic field $h$ on the forward (backward) current $j_{\mathrm{s}}^{+}$ ($j_{\mathrm{s}}^{-}$) and the rectification ratio $R$. In order to understand the behaviour of spin currents in finite-sized chains we resort to the language of spinons \cite{Faddeev1981, Haldane1983}, which are the elementary excitations in our system. These spinons carry $\sigma^{z}=\pm 1/2$ and will be referred to as the up- and down-spin spinons respectively. The external magnetic field has the same effect on the spinons as that of a chemical potential on electrons for which there are particle and hole like excitations. At high temperatures since all the spins are aligned in the up-direction [see fig.~\ref{fig:2}(b) top] the mobility of the down-spin spinon is much larger than the up-spin spinon causing the forward and backward currents to be negative [fig.~\ref{fig:3}(a) bottom panel]. Unfortunately, the rectification ratio $R$ [fig.~\ref{fig:3}(b) red dashed line] remains constant in this regime and is quite low, making high temperatures undesirable for tuning or observing $R$. In the low temperature regime since the two end spins are pinned (due to the bath) a non-uniform distribution of $\avl\sigma^{z}_{i}\avr$ is created in which the up-spins have a larger magnitude $|\avl\sigma^{z}\avr|$ as compared to the down-spins [see fig.~\ref{fig:2}(a) top]. This fact along with the competing effects of the anisotropy $\Lambda$ and the external magnetic field $h$ greatly affect the transport properties. At low external magnetic fields $h/\mathcal{J} < \Lambda$ the anisotropy (nearest-neighbor interaction) plays an important role to decide the lowest-energy spin excitation \cite{note1}, which is created by flipping an up-spin [for example the middle up-spin in fig.~\ref{fig:2}(a) top] to a down-spin, thus generating down-spin spinons. This causes the forward and backward currents to be negative as shown in fig.~\ref{fig:3}(a) top panel. At moderate external magnetic fields, i.e., when $h/\mathcal{J} > \Lambda$ but the system has an anti-ferromagnetic ordering, the external magnetic field  is more important than the anisotropy $\Lambda$. Thus, in this regime the lowest energy spin excitation is created by flipping one of the down-spins to an up-spin generating up-spin spinons, causing the spin currents to be positive. This change in sign of the spin current from negative to positive \cite{note2} is also reflected in the spin-Seebeck coefficient \cite{Furukawa2005}. In the regime of extremely high external magnetic fields, $h/\mathcal{J} \gg \Lambda$, all the spins will be aligned in the up-direction (Ferromagnetic ordering) causing the down-spin spinons to have a higher mobility and thus causing the spin-currents to be negative [not shown in Fig.~\ref{fig:3}]. Therefore, in order to avoid $R > 1$, in fig.~\ref{fig:3}(b) [blue solid line] we plot the rectification $R$ only at low fields, $h/\mathcal{J} < \Lambda$, so that only the down-spin spinons play a role in transport. Clearly the rectification can be easily tuned with the external magnetic field and it shows a variation from 0 to 0.5 which can be easily detected. Observing the top panel in fig.~\ref{fig:3}(a) and the blue solid line in fig.~\ref{fig:3}(b) we find that even though the rectification is largest at small magnetic fields the spin current is also the smallest in that regime. This might pose a problem in an experimental setup where the strength of the signal plays a crucial role and hence it is judicious to have a small but finite magnetic field so that a high signal to noise ratio is maintained to observe rectification. Thus, in small finite sized systems an extremely high value of rectification can be obtained in the low-temperature regime and it can be tuned using an external magnetic field whereas the effect diminishes rapidly as the temperature increases.
\section{Summary}
\label{sec:4}
We have investigated the transport of magnetization via nonitinerant spins in the one-dimensional $XXZ$ spin chain. In our model the baths were a set of harmonic oscillators containing \emph{no spins} and the system-bath interface generated a spin current due to the spin-Seebeck effect. The absence of spins from the baths made it impossible to use the traditional definition of spin current as $\drom\sigma_{\mathrm{\sm{L}}}^{z}/\drom t$ and hence we resorted to the local spin current definition using the lattice continuity equation. Since the spin current at the lowest order of perturbation is second order in the system-bath coupling, the nonequilibrium modified Redfield solution became an essential ingredient in order to capture the spin current accurately. 

Our model was then used to study spin-current rectification in the far from linear response regime. Two important aspects were examined namely, the effect of system size and the tunability of rectification ratio with the external magnetic field. In case of system size dependence we found that the rectification ratio shows an oscillatory behaviour depending on the number of spins present in the system. This unusual oscillatory effect was attributed to the presence of spin frustration in systems with even number of spins, which led to the lowering of rectification ratio $R$ as compared to the odd counterparts. This behaviour was found to be predominant only at low temperatures and at high temperatures, due to no preferential ordering of the spins, the effect completely disappeared. The low temperature regime played an important role even in the case of tuning the rectification ratio with an external magnetic field. At low temperatures the rectification ratio could be varied from 0 to 0.5, with the largest $R$ obtained at low magnetic fields. Thus, the oscillatory behaviour with system size and large variation of the rectification ratio by an external field allows one to engineer spin devices making magnetic insulators interesting candidates for spintronics.
\acknowledgements
The authors would like to thank Popkov Vladislav and Peter H\"{a}nggi for insightful discussions.


\begin{thebibliography}{0}
\bibitem{note1} In the low-temperature regime the lowest-energy spin excitation is sufficient to determine the carriers of transport and qualitatively explain the transport properties.
\bibitem{note2} One might speculate here that the change in sign from negative to positive is possibly related to the phase transition which occurs in the equilibrium $XXZ$ model from the N\'{e}el state to the XY state. Although more work is needed to confirm if such phase transitions exist in nonequilibrium as well.
\end{thebibliography}
\end{document}